\documentclass[pre,reprint,superscriptaddress,twoside]{revtex4-2}
\usepackage[version=3]{mhchem}
\usepackage{balance}
\usepackage{mathptmx}
\usepackage{graphicx} 
\usepackage{float}
\usepackage[english]{babel}
\usepackage{array}
\usepackage[T1]{fontenc}
\usepackage[usenames,dvipsnames]{xcolor}
\usepackage{hyperref}

\usepackage[separate-uncertainty = true,multi-part-units=single]{siunitx}
\usepackage{amssymb}

\definecolor{cream}{RGB}{222,217,201}

\begin{document}

\title{Granular aqueous suspensions with controlled inter-particular friction and adhesion}

\author{Lily Blaiset}
\affiliation{Universit\'{e} Paris Cit\'{e}, CNRS,  Mati\`{e}re et Syst\`{e}mes Complexes UMR 7057, F-75013, Paris, France.}
\altaffiliation{Laboratoire SIMM, ESPCI Paris, PSL University, CNRS, Sorbonne Universit\'{e}, 75231 C\'{e}dex 05Paris, France.}
\author{Bruno Bresson}
\affiliation{Laboratoire SIMM, ESPCI Paris, PSL University, CNRS, Sorbonne Universit\'{e}, 75231 C\'{e}dex 05Paris, France.}
\author{Ludovic Olanier}
\affiliation{Laboratoire SIMM, ESPCI Paris, PSL University, CNRS, Sorbonne Universit\'{e}, 75231 C\'{e}dex 05Paris, France.}
\author{\'Elisabeth Guazzelli}
\affiliation{Universit\'{e} Paris Cit\'{e}, CNRS,  Mati\`{e}re et Syst\`{e}mes Complexes UMR 7057, F-75013, Paris, France.}
\author{Matthieu Roch\'e}
\affiliation{Universit\'{e} Paris Cit\'{e}, CNRS,  Mati\`{e}re et Syst\`{e}mes Complexes UMR 7057, F-75013, Paris, France.}
\author{Nicolas Sanson}
\affiliation{Laboratoire SIMM, ESPCI Paris, PSL University, CNRS, Sorbonne Universit\'{e}, 75231 C\'{e}dex 05Paris, France.}

\begin{abstract}
    We present a simple route to obtain large quantities of suspensions of non-Brownian particles with \textit{stimuli}-responsive surface properties to study the relation between their flow and interparticle interactions. We perform an alkaline hydrolysis reaction on poly(methyl methacrylate) (PMMA) particles to obtain poly(sodium methacrylate) (PMAA-Na) particles. We characterize the quasi-static macroscopic frictional response of their aqueous suspensions using a rotating drum. The suspensions are frictionless when the particles are dispersed in pure water. We relate this state to the presence of electrosteric repulsion between the charged surfaces of the ionized PMAA-Na particles in water. Then we add monovalent and multivalent ions (Na$^+$, Ca$^{2+}$, La$^{3+}$) and we observe that the suspensions become frictional whatever the valency. For divalent and trivalent ions, the quasi-static avalanche angle $\theta_c$ at large ionic strength is greater than that of frictional PMMA particles in water, suggesting the presence of adhesion. Finally, a decrease in the pH of the suspending solution leads to a transition between a frictionless plateau and a frictional one. We perform Atomic Force Microscopy (AFM) to relate our macroscopic observations to the surface features of the particles. In particular, we show that the increase in friction in the presence of multivalent ions or under acidic conditions is driven by a nanoscopic phase separation and the bundling of polyelectrolyte chains at the surface of the particle. Our results highlight the importance of surface interactions in the rheology of granular suspensions. Our particles provide a simple, yet flexible platform to study frictional suspension flows.
\end{abstract}
\maketitle

\section{Introduction} Suspensions of particles dispersed in a liquid play a significant role in natural and industrial contexts. Examples of suspensions include sediments, biological fluids (\textit{e.g.} blood), and industrial products such as paints and inks. Understanding their mechanical behavior is crucial for efficient handling and reduction of environmental impact in processes such as water treatment, mineral processing, and material synthesis.

The history of suspension rheology dates back to the early 20th century with the seminal work of Einstein \cite{einstein_neue_1906, einstein_berichtigung_1911} on the effective suspension viscosity in the dilute regime. Subsequently, significant progress has been made in the semi-dilute regime.\cite{batchelor_hydrodynamic_1972, guazzelli_physical_2012} However, only recently has a description of concentrated suspension flows emerged that is in line with experimental observations. This conceptual advance acknowledges the growing importance of direct interparticle contacts over hydrodynamic interactions as the volume fraction of particles $\phi$ increases,\cite{guazzelli_rheology_2018} leading to a frictional description of the flow of suspensions inspired by that of dry granular media, which relates the pressure $P$ applied to the flowing suspension to the shear stress $\tau$ through a macroscopic friction coefficient $\mu = \tau/P$. \cite{forterre_flows_2008} Experiments support this approach. \cite{boyer_unifying_2011, guy_towards_2015, clavaud_revealing_2017, tapia_influence_2019, etcheverry_capillary-stress_2023} The importance of the contact contribution is especially seen close to the jamming transition, where the viscosity diverges at a critical volume fraction~$\phi_c$.\cite{lemaire_rheology_2023, gallier_rheology_2014} 

In this framework, the ability to tailor the details of interparticle interactions becomes crucial. Emerging research relies on engineering the surface properties of particles to impart specific rheological characteristics to the suspension. While many protocols exist for colloidal particles,\cite{ballesta_wall_2012, hsu_roughness-dependent_2018, hsu_exploring_2021} efforts in the non-colloidal regime of present interest are more recent. Control over interparticle interactions can be achieved by adjusting the roughness and/or friction of the particles on the microscopic scale \cite{tapia_influence_2019} or by controlling the nature of the interactions between the particles with adsorbed ions,\cite{clavaud_revealing_2017} surfactants,\cite{richards_turning_2020} or custom polymer coatings.\cite{guy_towards_2015, moratille_cross-linked_2022} However, these methods, especially the latter, produce small amounts of particles and require complex chemical protocols that are difficult to establish outside of specialized chemistry laboratories. To the best of our knowledge, a simple and scalable procedure to produce large quantities of particles with controllable surface properties once immersed in water with the aim of investigating the frictional behavior of non-Brownian suspensions is still missing. 

In this paper, we present a study of the surface properties and interactions of polyelectrolyte particles made of poly(sodium methacrylate) (PMAA-Na), obtained by complete alkaline hydrolysis of commercial particles made of poly(methyl methacrylate) (PMMA). PMAA-Na particles can be mass-produced using this technique, and their dispersion in water is stable. We perform rotating drum experiments to characterize the quasi-static macroscopic friction coefficient $\mu_c$ of aqueous suspensions of PMAA-Na particles in the dense regime through a measurement of the quasi-static avalanche angle $\theta_c$. First, we observe that these permeable particles are frictionless in pure water compared to bare PMMA particles. The addition of monovalent salt in solution leads to an increase in the friction coefficient. We take advantage of the polyelectrolytic nature of our particles to modulate their frictional behavior by tuning the ion valency, pH, or temperature. We discuss our macroscopic observations in light of a microscopic characterization of the particle surface using atomic force microscopy.

\section{Material and Methods} 

\subsection{Alkaline hydrolysis of PMMA particles} 
\begin{figure} 
\centering
\includegraphics[width=\linewidth]{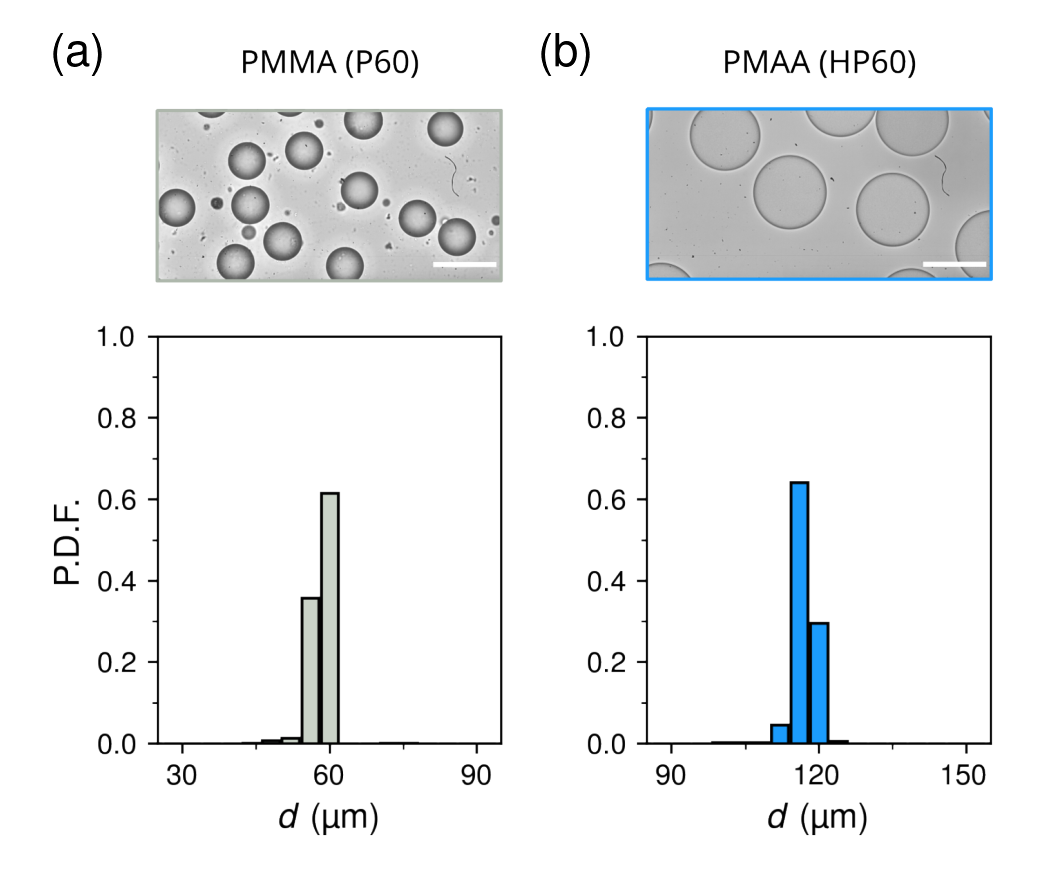}
\caption{Chemical modification of hydrophobic PMMA particles (P60) into hydrophilic PMAA-Na particles (HP60). Microscopic images and size distributions of P60 (a) and HP60 (b) particles. (Scale bars 100 µm)}
\label{fig-1}
\end{figure}

We start with cross-linked monodisperse (3 mol\%) poly(methyl methacrylate) particles with diameter $d_{P60} = \SI{60(2)}{\micro\meter}$ (P60) and density $\rho_{P60} = \SI{1169}{\kilo\gram\per\meter\cubed}$ (CA-60 Spheromers, Microbeads\textsuperscript{\tiny\textregistered}) (see Figure \ref{fig-1}a). These particles are synthesized in the presence of a steric stabilizer, polyvinylpyrrolidone (PVP), to make them water-dispersable.
We perform alkaline hydrolysis on these particles following a protocol inspired by Thanoo and Jayakrishnan.\cite{thanoo_preparation_1990} We disperse 10~g of P60 particles in 240 ml of ethylene glycol (Sigma-Aldrich) containing 25.75 g of potassium hydroxide (KOH pellets, Sigma-Aldrich, $\geq 85$ wt\%). Reactants and solvents are used as received. The reaction medium is kept at a controlled temperature of \SI{160}{\degreeCelsius} for 6 hours. P60 particles are then completely hydrolyzed (see Figure \ref{fig-1}b). This reaction can be easily scaled up to 50 g. We find that potassium hydroxide (KOH) and methacrylate groups (-CH$_3$COOCH$_3$, MMA) must be in a ratio of 4 KOH to 1 MMA group for the reaction to be optimal. These groups react through an addition-elimination mechanism,\cite{azevedo_alkaline_2020} resulting in hydrophilic poly(potassium methacrylate) particles and methanol as a by-product. We purify the modified particles with a series of 5min-centrifugation at 2880 \textit{g}: once to remove the supernatant, then twice in a saturated solution of sodium hydrogencarbonate (NaHCO$_3$, Sigma-Aldrich, pH = 8.3) to remove excess KOH. Finally, the particles are washed with ultrapure water (final~pH~=~7, conductivity $\sigma \leq \SI{100}{\micro\siemens\per\centi\meter}$) using a 106-micron sieve. At the end of the reaction, we collect hydrolyzed particles made of poly(sodium methacrylate) (PMAA-Na) (HP60). Particles are freeze-dried overnight and then stored at room temperature. Figure \ref{fig-1} shows that the profile density functions (P.D.F.) related to the size distributions of P60 and water-immersed HP60 particles are narrow, \textit{i.e.} these particles are monodisperse. Freeze-dried HP60 particles have a diameter $d_{HP60, dry} = \SI{59(2)}{\micro\meter}$ and a density $\rho_{HP60,dry} = \SI{1373}{\kilo\gram\per\meter\cubed}$. They swell to an average diameter $d_{HP60, wet} = \SI{118(2)}{\micro\meter}$ when immersed in pure water, with a density $\rho_{HP60, wet} = \SI{1041}{\kilo\gram\per\meter\cubed}$ and a swelling ratio of $d_{HP60, dry}/d_{HP60, wet} \approx$ 2. Indirect titration shows that 65 mol\% of methylmethacrylate (MMA) groups are converted into sodium methacrylate (see Figure S1 \textit{Supporting Information}).

\subsection{Suspending liquids} We test the dependence of interparticle interactions in suspension on physicochemical conditions by dispersing P60 and HP60 particles in aqueous electrolyte solutions. We investigate the effect of ionic strength $I = \frac{1}{2} \sum c_iz_i^2$, where $c_i$ is the concentration of ion $i$ and $z_i$ its valency, and ion valency with sodium chloride (NaCl, Sigma-Aldrich), calcium chloride (CaCl$_2$, purchased in its hydrated form CaCl$_2 \cdot$6H$_2$O from Sigma-Aldrich), and lanthanum chloride (LaCl$_3$, purchased in its hydrated form LaCl$_3 \cdot$7H$_2$O from Sigma-Aldrich). We choose these salts because they are mild Lewis acids and have little effect on the pH of the solution. We vary the electrolyte concentration between $5\cdot10^{-5}$ and 0.5 M, with a pH = 6-7. We test the influence of acidity by varying the pH in acidic solutions of hydrochloric acid (HCl, Sigma-Aldrich). HP60 particle suspensions are prepared in water, and the pH is adjusted from 7 to 2. All chemicals are used as received.

\subsection{Rotating drum} Conventional rheological methods typically analyze suspension flows under constant volume constraints. However, these methods do not offer insight into the macroscopic friction coefficient of the suspension, a crucial parameter that we aim to explore in this paper. Recently, alternative approaches have emerged to characterize dense suspensions, drawing inspiration from pressure-based experiments conducted with dry granular flows.\cite{boyer_unifying_2011, clavaud_revealing_2017, etcheverry_capillary-stress_2023}

\begin{figure} 
\centering
\includegraphics[width=\linewidth]{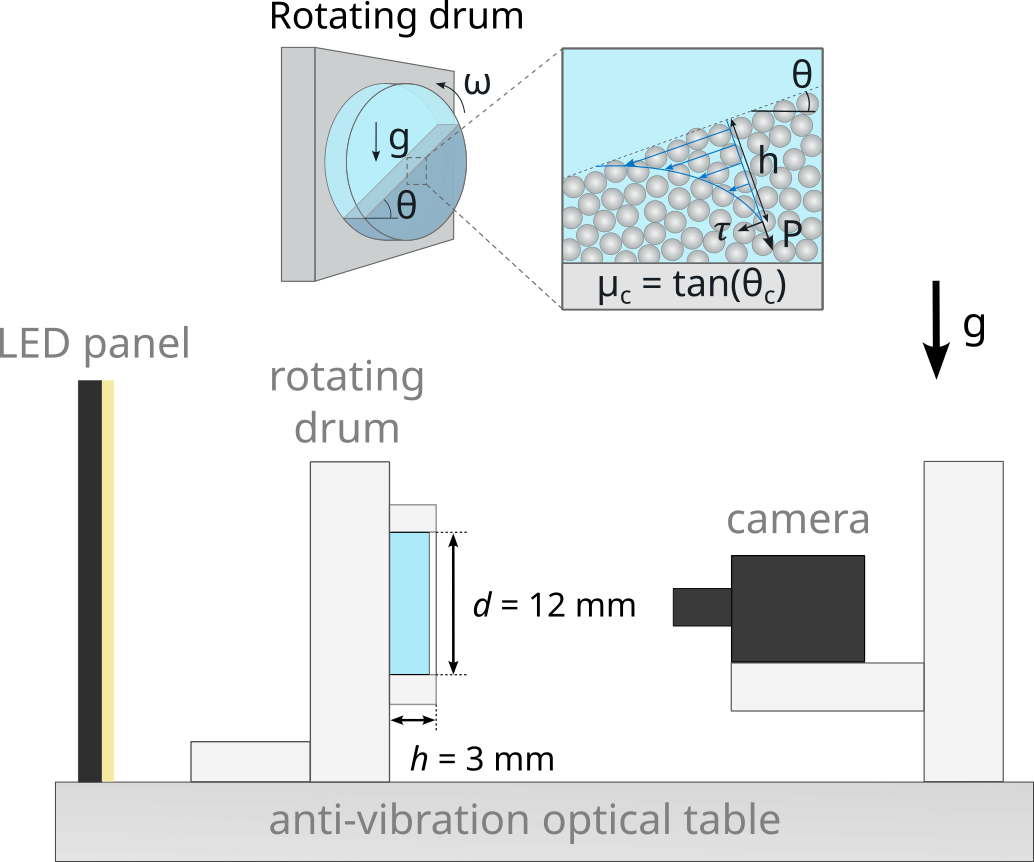}
\caption{Experimental setup used for rotating drum experiments. The macroscopic friction coefficient $\mu$ of the suspension is defined by $\mu = \tau/P$, with $\tau$ the tangential shear stress and $P$ the confining pressure applied on the flowing layer of immersed particles. The quasi-static macroscopic coefficient $\mu_c$ is directly related to the quasi-static avalanche angle ($\theta_c$) of the suspension as $\mu_c~=~\tan{\theta_c}$.}
\label{fig-2}
\end{figure}

We take inspiration from Clavaud \textit{et al.}\cite{clavaud_revealing_2017, clavaud_suspensions_2018} to build a rotating drum. This setup is a convenient way to obtain information on the quasi-static macroscopic friction coefficient of an assembly of particles.\cite{courrech_du_pont_granular_2003} Our drum has a diameter $d$ = 12 mm and a thickness $h$~=~3~mm. The complete setup used for rotating drum experiments is sketched in Figure \ref{fig-2}. Particles immersed in water at different salt concentrations or pH are introduced into the drum. After sealing, the drum is mounted on a precision rotating stage (M-061.PD from Physik Instrumente). Measurements are performed at room temperature, unless otherwise specified. In this configuration, the macroscopic friction coefficient $\mu$ of the flowing layer of particles ($\sim 10d$) depends on the tangential stress $\tau$ and the confining pressure $P$ of the configuration, such as $\mu = \tau/P$.

After verifying the limit of the quasi-static regime (see Figure S2 in \textit{Supporting Information}), a speed of $\omega = \SI{0.1}{\degree\per\second}$ is chosen to perform all experiments. Before each measurement, the particles are resuspended for 5 min at \SI{90}{\degree\per\second}, and then left to sediment for 5 min. To start the measurements, we impose a rotational speed $\omega = \SI{0.1}{\degree\per\second}$ for 1 h. We record images with a Basler camera and a macro lens at a frame rate ranging from 0.0056 to 0.0083 images.\si{\per\second}. Image processing is done with a custom Python code. We also built a temperature-controlled drum to perform experiments at various temperatures ranging from 10 to 60~°C (see setup in Figure S3 in \textit{Supporting Information}). Temperature is set by a thermostated bath. We wait 20 min for the desired temperature to stabilize before starting the measurements. We checked that we do not observe stick-slip avalanches in our experiments, \textcolor{black}{meaning that the maximum avalanche angle $\theta_m$ reached as the slope of the pile of particles increases and the angle of repose $\theta_r$ where the particles start to avalanche are equivalent}.\cite{nowak_maximum_2005} By fixing the rotational speed at \SI{0.1}{\degree\per\second}, we ensure that experiments are carried out under quasi-static conditions, in a continuous avalanche regime (rolling regime), where $\theta_m = \theta_r$. Therefore, we define $\theta_c$ as the quasi-static avalanche angle in our measurements. The quasi-static macroscopic friction coefficient is given by $\mu_c = \tan{\theta_c}$. Finally, we compute the particle Stokes number,\cite{clavaud_revealing_2017} 
\begin{equation}
    St = \frac{\sqrt{\rho_p\Delta\rho g d^3}}{18\eta_f},
\end{equation}
and the particle-to-fluid density ratio $r = (\rho_p/\rho_f)^{1/2}$,\cite{courrech_du_pont_granular_2003} where $\Delta\rho_p = \rho_p - \rho_f$, $g$ being the gravity, $d$ the particle diameter, and $\eta_f$ the fluid viscosity. In both P60 and HP60 suspensions, $r \approx 1.1$ and $St \approx 10^{-2}$, meaning that inertial effects are negligible. 

\subsection{AFM surface scanning} We use atomic force microscopy (AFM) to characterize the surface of our particles. Dry particles (P60 and HP60) are deposited on a glass substrate covered with a thin layer of Cyanolit\textsuperscript{\tiny\textregistered} glue. The sample is dried overnight, then carefully rinsed with Milli-Q water to remove particles that are not glued to the substrate. A few water drops are placed on the substrate before putting the sample under the AFM apparatus. A Bruker AFM (Dimension Icon\textsuperscript{\tiny\textregistered}) configured in Peak Force Quantitative Nanomechanical (QNM) tapping mode is used to image the surface at the particle apex in aqueous media. Each image pixel contains a force curve that we can relate to topographical features. We obtain the latter with ScanAsyst-Fluid probes (Bruker, tip radius $r \approx$ 20 nm, frequency $f~\approx$~150~kHz, spring constant $k \approx$ \SI{0.7}{\newton\per\meter}). We perform a force-distance curve on a glass plate to measure the beam deflection at the diode output in volts.

\subsubsection{Topography.~~} To probe surface roughness, we change the scan size to 10x5 or 5x2.5 \si{\micro\meter\squared}, while the scanning rate, the normal force and the peak force amplitude are set at 0.3 Hz, 350 pN, and 300 nm, respectively. The application of a low normal force is crucial to prevent the cantilever from damaging the particle surface, and the high amplitude prevents the cantilever from damaging or deforming the surface. 

We acquire a series of images of the apex of the same particle to observe the effects of ionic strength $I$ with NaCl and ion valency with CaCl$_2$ and LaCl$_3$. We increase the salt concentration \textit{in-situ}, \textit{i.e.} during an experiment by pumping a volume of suspending fluid out with a syringe and replacing it with a few drops of the new salt solution at the target concentration. With this protocol, the sample always remains immersed. We repeat this process three to four times (10 to 15 min) until the sample reaches equilibrium, \textit{i.e.} when the diameter of the particles on the substrate decreases until it reaches a constant value. We apply the same protocol for pH variation with HCl solutions.

We process the scans with the NanoScope Analysis software. We apply a second-order flattening to the topography image and estimate the surface roughness as the root mean square average $R_q$ of vertical displacements from the mean image data plane:
\begin{equation}
    R_q = \sqrt{\frac{\sum(Z_i)^2}{N}},
\end{equation}
where $Z_i$ is the local displacement and N is the number of points within the selected region of interest. The root mean square $R_q$ is commonly reported to calculate surface roughness, due to its high sensitivity to tip-sample distortion.


\subsubsection{Young modulus measurements.~~} The Young modulus of HP60 particles in pure water is determined from images with a scan size ranging between 0.25 and \SI{25}{\micro\meter\squared}, with a normal force from 1.5 to 2.5 nN. We extract an estimate of the Young modulus using a standard procedure of the Nanoscope Analysis software, which automatically fits the force-distance curves of the image with a Derjaguin-Muller-Toropov (DMT) model. We obtain an average Young modulus $E = \SI{500}{\kilo\pascal}$. \textcolor{black}{However, because we focused on surface topography, the exact tip stiffness usually measured by acquiring a thermal noise spectrum has not been accurately measured. The procedure gives us an order of magnitude of the Young modulus of HP60 particles. Young moduli values of poly(methyl methacrylate) scale around 3.5 GPa.\textcolor{black}{\cite{ishiyama_effects_2002}} The order of magnitude of the Young modulus determined for HP60 particles under water is reduced to several hundred kPa. This lower Young modulus is in agreement with the swelling of HP60 particles in water, giving less mechanical resistance compared to P60 particles.}

\section{Results and discussion}

\subsection{Hydrolyzed particles are frictionless in pure water}

\begin{figure}
\centering
\includegraphics[width=\linewidth]{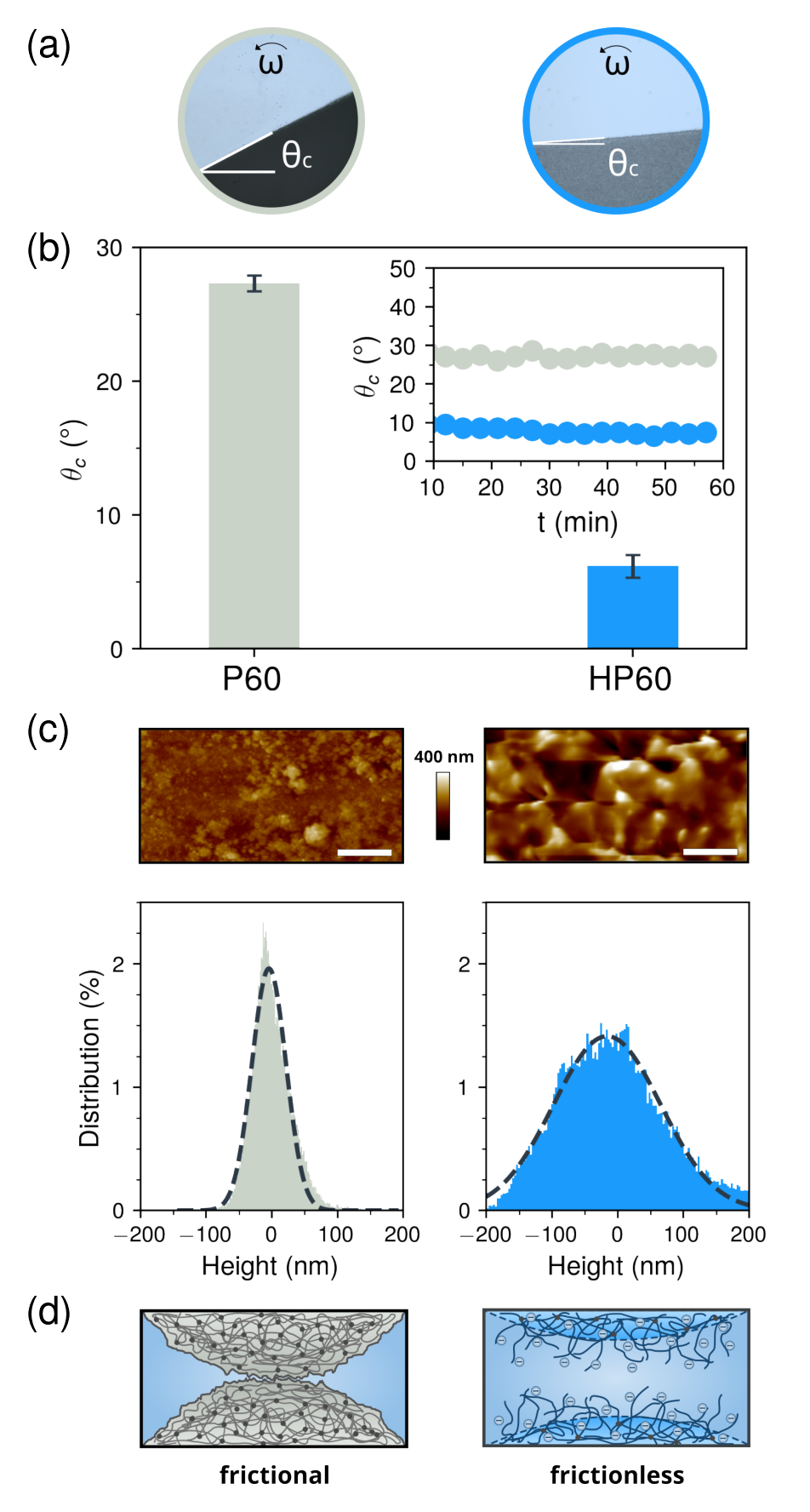}
\caption{Frictional versus frictionless behavior of P60 (left) and HP60 (right) particles suspended in pure water (pH = 7). (a) Rotating drum images (b) Average quasi-static avalanche angle $\theta_c$ for P60 and HP60 particles. Inset: Avalanche angle $\theta_c$ versus time. (c) AFM images, roughness profile height, and corresponding Gaussian distribution. (d) Schematics of P60 show particle-particle contact and HP60 shows lubrication at the interface due to electrosteric repulsion in pure water.}
\label{fig-3}
\end{figure}

As shown in Figure \ref{fig-3}a-b (left), steady avalanches of P60 particles reach an angle $\theta_c = \SI{27.3(0.6)}{\degree}$, giving a quasi-static macroscopic friction coefficient $\mu_c = 0.52$, which are typical values expected for standard frictional granular materials and suspensions.\cite{courrech_du_pont_granular_2003, clavaud_revealing_2017, carrigy_experiments_1970} 
However, this value is relatively high compared to what has been observed with other experimental setups, such as pressure-imposed rheometers. In this type of configuration, the quasi-static macroscopic friction coefficient $\mu_c$ of a frictional suspension is generally lower, with $\mu_c \approx$ 0.3-0.4. \cite{tapia_influence_2019, boyer_unifying_2011, etcheverry_capillary-stress_2023} We explain this difference by the different confining pressures applied to the particles. In pressure-imposed rheometers, particles are subjected to pressure typically on the order of a few tens of Pascals, while in the rotating drum configuration, the confining pressure generated by the fluid imposed on the flowing particles can be calculated as:\cite{perrin_interparticle_2019}
\begin{equation}
    P = \phi \Delta \rho gh \cos(\theta_c),
\end{equation}
with $\phi$ the volume fraction of the suspension, $\Delta \rho = \rho_{HP60} - \rho_f$ the density difference between the particles and the suspending fluid at \SI{25}{\degreeCelsius}, $g$ the acceleration of gravity, $\theta_c$ the avalanche angle of the suspension, and $h$ the particle layer flowing in the steady-state regime. We estimate the thickness of the flowing particle layer of the particles to be $h \approx 10d$, where $d$ is the particle diameter. 
In the case of P60 particles, the pressure acting on the flowing layer of the particles is $P \sim \SI{0.5}{\pascal}$. With HP60 particles, $P \sim \SI{0.3}{\pascal}$. \textcolor{black}{The confinement pressure can influence particle contacts and interactions, as the inter-particle friction can depend on the particle pressure. At low pressures, friction is mainly controlled by adhesion because most of the friction force comes from the energy needed for surfaces to detach and move over each other.\cite{israelachvili_intermolecular_2011} Therefore, the low pressure applied to both P60 and HP60 particles makes their avalanche angles sensitive to particle contacts and interactions.} For P60 particles, \textcolor{black}{the quasi-static avalanche angle $\theta_c$ and the quasi-static macroscopic friction coefficient $\mu_c$ are considerably larger than those of typical frictional spheres,\cite{clavaud_revealing_2017} suggesting the possible presence of weak cohesive effects. These interactions, only be detectable under the low pressure imposed on the particles, can be due to the presence of a steric stabilizer (PVP) 
used during the manufacturing process of the particles.}

Much lower values of the avalanche angle are obtained for hydrolyzed HP60 particles (Figure \ref{fig-3} a-b), $\theta_c = \SI{6.2(0.8)}{\degree}$. This value corresponds to a quasi-static macroscopic friction coefficient $\mu_c = 0.11$, typical of a frictionless suspension.\cite{clavaud_revealing_2017, peyneau_frictionless_2008} We may then wonder if and how the hydrolysis reaction affects the surface roughness $R_q$ of the particle. We characterize the surface topography of P60 and HP60 particles with AFM scans from which we extract the surface height distribution (Figure \ref{fig-3}c). We see that the surface height distribution of both types of particles is well-fitted with a Gaussian function. The height distribution of \textcolor{black}{frictionless} HP60 particles (Figure \ref{fig-3} d) is broader than that of \textcolor{black}{frictional} P60 particles. HP60 particles have a roughness $R_q$ = 70 $\pm$ 16 nm, larger than $R_q$~=~26~$\pm$~6~nm for P60 particles. The greater surface contrast and roughness we measure characterize the swollen PMAA-Na network immersed in water. If the particles had no affinity for the solvent and were in direct contact,\cite{pohlman_surface_2006} HP60 particles should display a larger $\theta_c$ than their non-hydrolyzed counterparts, in contrast to our results showing the opposite. We think that the smaller quasi-static friction coefficient $\mu_c$ of HP60 particles compared to P60 particles results from the fact that the former are made of PMAA-Na, a polyelectrolyte that will induce electrosteric repulsion between the particles.\cite{kagata_surface_2001, ohsedo_surface_2004} This interaction, which is in general a combination of an electrostatic repulsion and a steric one due to the impossibility of the polymer networks of two particles to interpenetrate, screens roughness effects and permits lubrication of the contact (Figure \ref{fig-3}d). In contrast, P60 particles show no electrosteric repulsion and are in direct contact. The nature of this repulsion opens vast possibilities for controlling it with various physicochemical \textit{stimuli} that we explore in the rest of the paper.

\subsection{The ionic strength and valence of ions controls friction and adhesion}

The material our particles are made of, PMAA-Na, is a polyelectrolyte in water that is responsive to ionic strength and ion valency.\cite{li_effects_2022, drifford_physical_2001} We investigate these effects using sodium chloride (NaCl), calcium chloride (CaCl$_2$) and lanthanum chloride (LaCl$_3$). Molar concentrations of monovalent, divalent, and trivalent salts are calculated such as the ionic strength varies $I$ from 5.10$^{-5}$ to 0.5 M. First, optical microscopy indicates that the particle size decreases as the ionic strength increases (Figure S4 in \textit{Supporting Information}). The decrease in particle size with an increase in ionic strength results from osmotic effects between the ions confined within the particle network and those in the surrounding bulk solution.\cite{eichenbaum_ph_1998} Second, we observe that the suspensions become more frictional as $I$ increases. At low ionic strength ($I \leq$ 5.10$^{-3}$ M), the quasi-static avalanche angle increases with the ionic strength, independent of both the nature of the ion and its valency (Figure \ref{fig-4} \textcolor{black}{and Figure S5 in \textit{Supporting Information}).} Hence, at these concentrations, load and decreasing solvation of the polyelectrolyte chains drive macroscopic friction.\cite{israelachvili_intermolecular_2011} However, for $I_c$ $>$ 5.10$^{-2}$ M, the avalanche angle $\theta_c$ of HP60 particles appears to saturate in the presence of NaCl while it still increases in the presence of Ca$^{2+}$ or La$^{3+}$, even exceeding the value of $\theta_c$ obtained for a suspension of frictional P60 particles (Figure \ref{fig-4}). The effect is more pronounced for La$^{3+}$ than for Ca$^{2+}$. Thus, not only do multivalent ions induce a larger apparent friction coefficient than monovalent ones, but friction coefficients also increase with valency.


\definecolor{peach}{RGB}{253,114,114} 
\definecolor{honey}{RGB}{234,181,67}
\definecolor{keppel}{RGB}{88,177,159}

\begin{figure} 
\centering
\includegraphics[width=\linewidth]{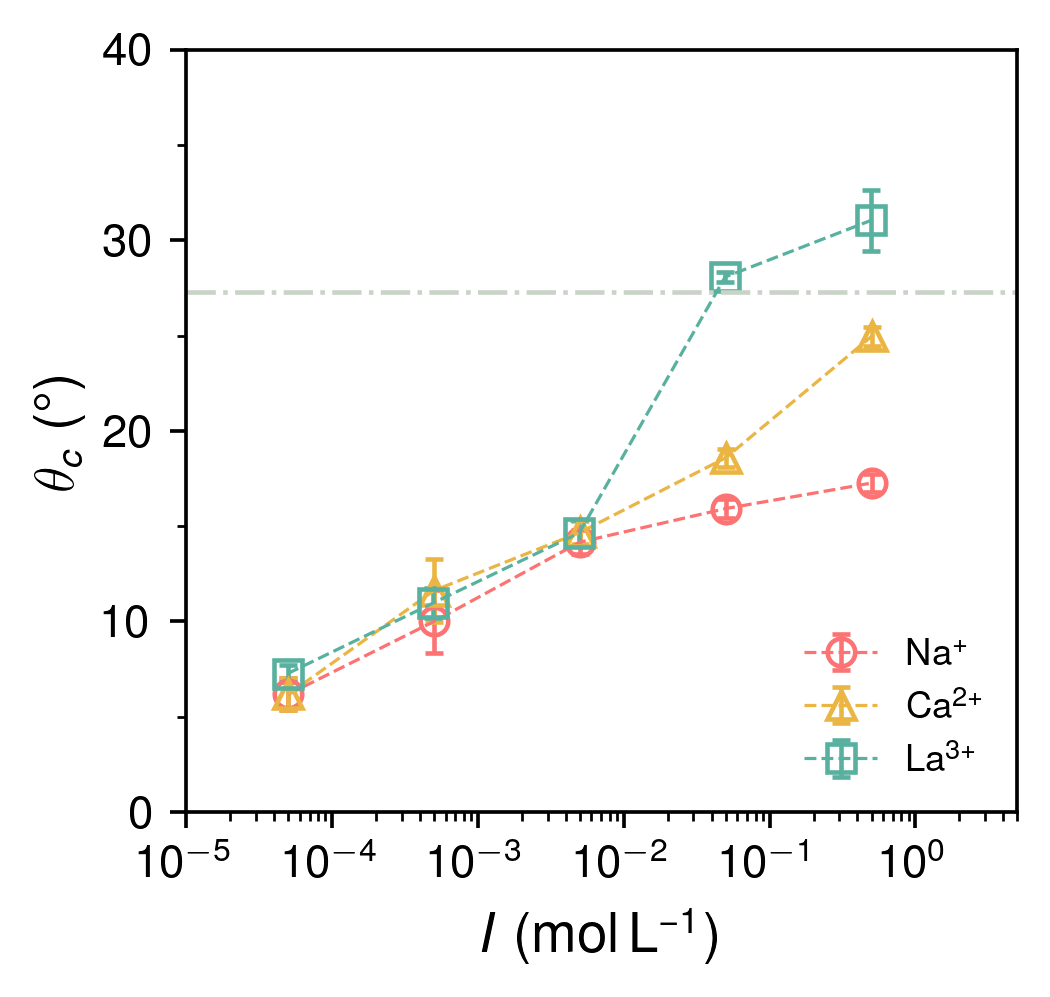}
\caption{Impact of ion valency on the frictional properties of HP60 particles. Quasi-static avalanche angle $\theta_c$ of HP60 suspensions with monovalent cation Na$^+$ (\textcolor{peach}{$\bigcirc$}), divalent cation Ca$^{2+}$ (\textcolor{honey}{$\triangle$}), and trivalent cation La$^{3+}$ (\textcolor{keppel}{$\square$}) with increasing ionic strength $I$. The grey dashed line corresponds to the avalanche angle $\theta_c$ of PMMA particles P60.}
\label{fig-4}
\end{figure}

\begin{figure} 
\centering
\includegraphics[width=\linewidth]{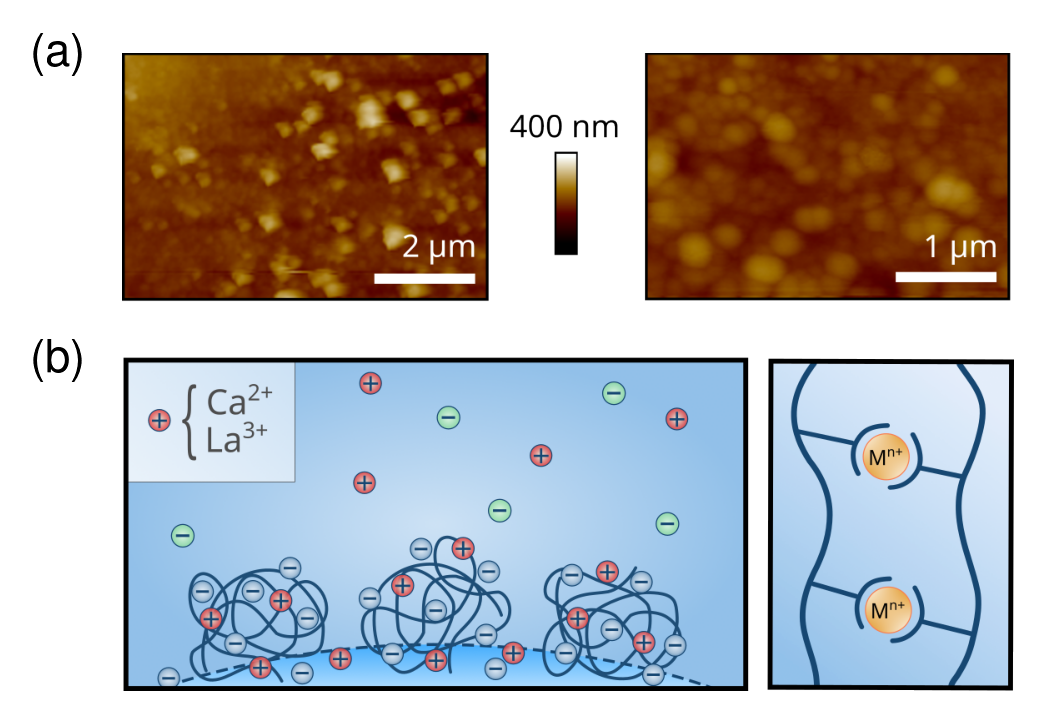}
\caption{Surface patterning resulting from phase separation on HP60 particles with increasing ion valency. (a) AFM images of HP60 particles with CaCl$_2$ = 0.17 M (left) and LaCl$_3$ = 0.11 M (right). (b) Schematic of the particle surface HP60 in the presence of multivalent cations Ca$^{2+}$ or La$^{3+}$.}
\label{fig-5}
\end{figure}

Direct analysis of the force-displacement images obtained with the AFM provides interesting clues on the dependence of the particle surface profile as a function of the valency of the ions. For example, we observe inhomogeneities on HP60 particles in the presence of multivalent cations, as shown in Figure \ref{fig-5}a. These nodules have a lateral size $l$ = \SI{284(114)}{\nano\meter} and $l$ = \SI{61(20)}{\nano\meter} for HP60 particles in presence of CaCl$_2$ and LaCl$_3$, respectively. This observation is reminiscent of the formation of bidentate and tridentate complexes between charged groups and cations observed on polyelectrolyte brushes in the presence of multivalent cations.\cite{williams_grafted_1993, yu_multivalent_2018} In this case, the formation of nodules is attributed to electrostatic bridging between polyelectrolyte chains inside the brush, leading to brush collapse. As illustrated in Figure \ref{fig-5}b, polyelectrolyte chains at the periphery of the particle can adopt a pinned micelle-like structure in the presence of divalent or trivalent cations. The pinned micelle-like structure seems to be accentuated in the case of trivalent cations (La$^{3+}$), since the addition of trivalent cations emphasizes the lack of polymer solvation.\cite{williams_grafted_1993} This is consistent with the reduced swelling behavior of HP60 particles with La$^{3+}$ (see Figure S4 in \textit{Supporting Information}). By choosing water as a suspending fluid to study HP60 particles, which is typically a good solvent for polyelectrolyte chains in the absence of added salt, we can differentiate between the effects of low solubility and the influence of cation valency on surface structure. When the chains bundle, they also become neutral and lose affinity for water. A hydrophobic effect may appear on an interparticular scale that tends to induce adhesive forces between the particles.\cite{israelachvili_intermolecular_2011, yu_multivalent_2017} These additional adhesive forces in the presence of multivalent cations could induce greater macroscopic friction in those suspensions compared to those of monovalent cations.

\subsection{Effect of pH and temperature}

\begin{figure} 
\centering
\includegraphics[width=\linewidth]{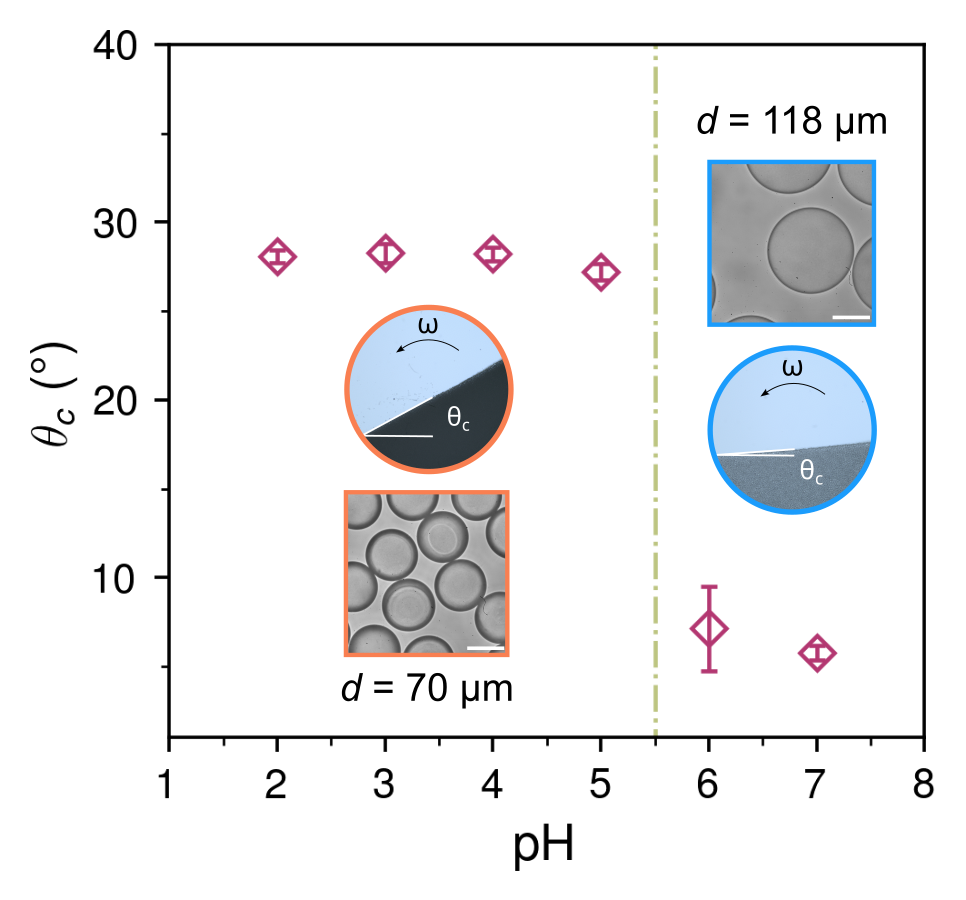}
\caption{Quasi-static avalanche angle $\theta_c$ and surface roughness $R_q$ of HP60 particles showing pH dependence. Inset: microscopic and drum images at low pH and high pH (Scale bars 50 µm). At pH $\leq$ 5, HP60 particles are in their protonated form which induces H-bonding and hydrophobic interactions, resulting in a phase separation of the polymer brushes at the particle surface. In acidic conditions, the avalanche angle reaches a plateau at $\theta_c$ = \SI{28.2(0.6)}{\degree}.}
\label{fig-6}
\end{figure}

\begin{figure*} 
\centering
\includegraphics[width=\linewidth]{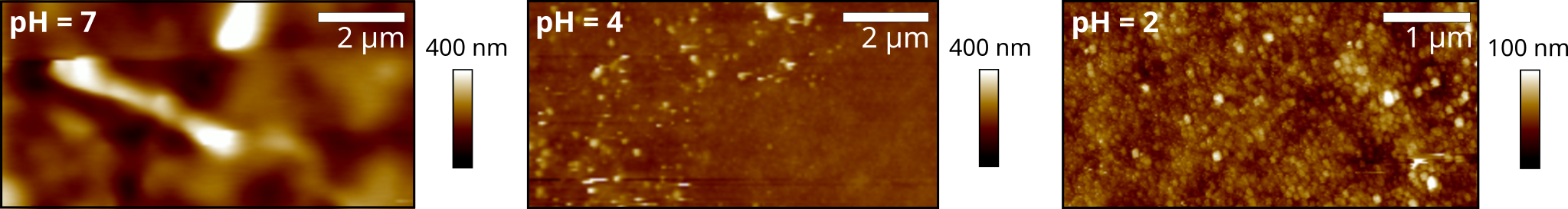}
\caption{Evolution of an HP60 particle surface in acidic conditions. Series of AFM images of a hydrolyzed particle HP60 in water with pH adjusted at 7, 4, and 2.}
\label{fig-7}
\end{figure*}

 We also investigate the effect of the pH of the solution on friction by adding hydrochloric acid (HCl) to the suspending liquid. In pure water (pH = 7), we observe that HP60 particles are frictionless, with an avalanche angle $\theta_c = \SI{6.2(0.8)}{\degree}$ (Figure \ref{fig-6}), in agreement with the data in Figure \ref{fig-3}. A decrease in pH shows a transition between two plateaus in the swelling and frictional behavior of HP60 particles. Above pH $\approx 5.5$, \textit{i.e.} in mild basic conditions, the particle size is not affected by the pH ($d \approx \SI{118}{\micro\meter}$) and the quasi-static avalanche angle ($\theta_c$) remains at relatively low values ($\theta_c \leq \SI{8}{\degree}$). Below pH = 5, the particles shrink by a factor close to 2, with $d = \SI{70(1)}{\micro\meter}$. At the same time, the values $\theta_c$ increase sharply, reaching a plateau at \SI{28.2(0.1)}{\degree}, typical of a macroscopic frictional behavior. The reduced swelling and increased friction observed for HP60 particles at low pH can be attributed to the protonation of the methacrylate groups (-COO$^-$), leading to the formation of methacrylic acid groups (-COOH). The protonated polymer presents a more hydrophobic character in comparison to its ionic version.\cite{jha_ph_2014} Adjusting the pH results in a sharper transition from frictionless to frictional behavior compared to adding salts. The change in the acid dissociation constant of PMAA (pKa~$\sim$~5.1)\cite{merle_synthetic_1987} indicates a transition from predominantly ionized groups to predominantly non-ionized groups.
 
We report in Figure \ref{fig-7} AFM images and surface roughness values $R_q$ determined for an apex of a particle at pH 7 (water), 4, and 2. While in pure water (pH = 7) poly(sodium methacrylate) (PMAA-Na) chains are extended in water, we observe the formation of nodules of non-ionized poly(methacrylic acid) (PMAA) chains at the particle surface under acidic conditions (pH = 2). Here, the nodules have a lateral size of $l$ = \SI{24(6)}{\nano\meter}. Due to hydrogen bonding, poly(methacrylic acid) (PMAA) chains collapse and undergo a conformation change, \textit{i.e.} a phase separation of PMAA at the particle surface. This is consistent with previous work on the swelling behavior of PMAA microgel particles.\cite{saunders_polymethyl_1997} This phase separation may induce adhesive forces of hydrophobic origin between particles, leading to high macroscopic friction coefficients in suspension.\cite{hadziioannou_forces_1986, israelachvili_intermolecular_2011}

\begin{figure} 
\centering
\includegraphics[width=\linewidth]{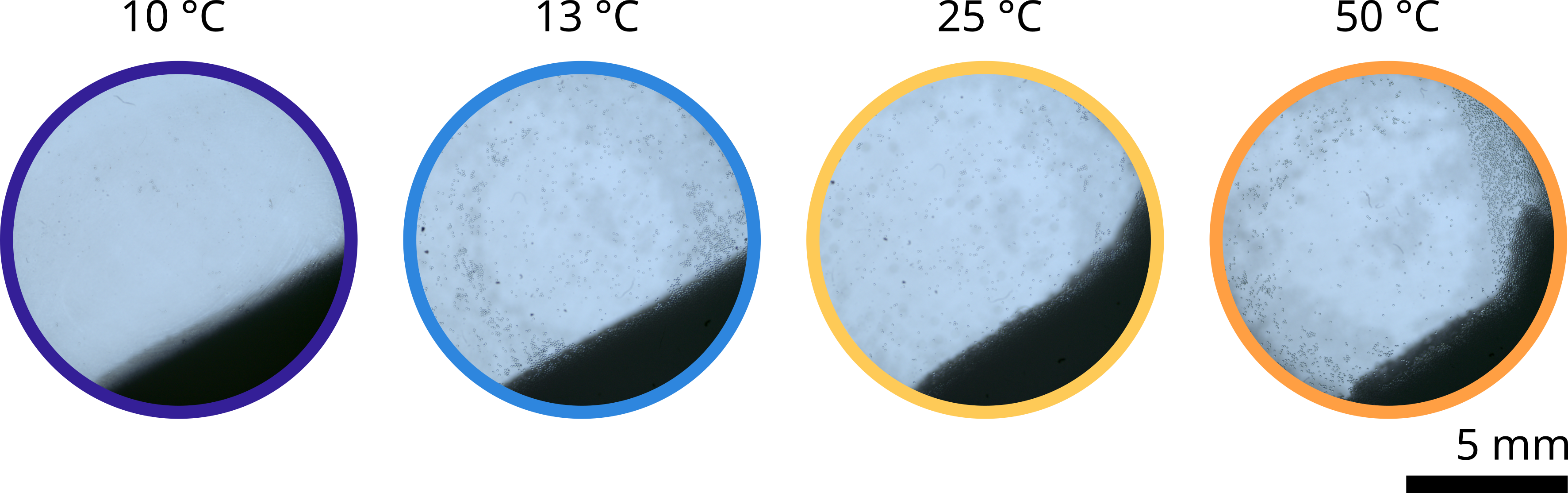}
\caption{Effect of temperature on hydrolyzed particles (HP60) under acidic conditions (pH = 4). From left to right: the temperature (T) varies from 10, 13, 25 to 50 °C. The avalanche angle ($\theta_c$) increases due to the accentuation of phase separation and hydrophobic effects.}
\label{fig-8}
\end{figure}

Finally, poly(methacrylic acid) (PMAA) exhibits a lower critical solution temperature (LCST) behavior under acidic conditions,\cite{robin_unexpected_2022}  \textit{i.e.} PMAA chains become insoluble and precipitate out of solution above the LCST. We assess the effect of temperature on the frictional behavior of protonated particles by measuring the avalanche angle $\theta_c$ of HP60 particles under acidic conditions (pH = 4) at temperatures of 10, 13, 25, 50, and \SI{60}{\degreeCelsius} (Figure \ref{fig-8} and video in \textit{Supporting Information}). With increasing temperature, phase separation and hydrophobic interactions due to methyl groups (-CH$_3$) are emphasized in the rotating drum, resulting in a hindered flow of HP60 particles under acidic conditions at high temperature (\SI{60}{\degreeCelsius}) likely resulting from strong adhesive forces between particles.

\section{Concluding remarks}

The major outcome of this work is the design of model non-Brownian suspensions of spheres with \textit{stimuli}-responsive surface properties that offer control of friction and possibly adhesion between the particles. The results we obtain in the rotating drum show that parameters such as the ionic strength, ion valency, pH in solution, or temperature are reliable proxies for adjusting the flow properties of suspensions of our polyelectrolyte particles. 

When first comparing the P60 particles with their hydrolyzed form HP60 in pure water, we find macroscopic friction coefficients typical of a frictional suspension for P60 particles ($\mu_c$~=~0.52) in contrast to HP60 particles which appear to be frictionless ($\mu_c$~=~0.11). We think that this frictionless state stems from an electrosteric repulsion between polyelectrolyte particles. Polyelectrolytes being known to be \textit{stimuli-}responsive, we tune the macroscopic frictional properties of HP60 particles using physicochemical parameters such as ionic strength, ion valency, pH, and temperature. By adding monovalent cations Na$^+$ to the medium, osmotic effects induce the collapse of polymer chains. The absence of electrostatic repulsion causes direct contact between particles, increasing interparticle friction. Using multivalent cations (Ca$^{2+}$ and La$^{3+}$), we observe a significant increase in the avalanche angle $\theta_c$ above a critical value of the ionic strength ($I_c = 10^{-2}$ M). This phenomenon arises from complex formation at the particle surface, causing phase separation between the polymer network and the solvent, leading to adhesive forces between particles and increased frictional properties in suspension. Similarly, protonation of HP60 particles under acidic conditions induces phase separation and adhesive forces, particularly at temperatures above~50~°C, where the suspension ceases to flow.

The microscopic surface features of HP60 particles immersed in saline solutions suggest interesting directions for future studies. First, the formation of nodules in the presence of multivalent ions and under acidic conditions point towards the occurrence of phase separation at the surface of the particles. We already discussed the possible existence of adhesion in our systems because of this phase separation, suggested by the large value of the avalanche angles with Ca$^{2+}$ and La$^{3+}$ and the formation of lumps at high temperature under acidic conditions. It would be interesting to explore how mixtures of monovalent and trivalent ions affect the interaction of particles, following recent results on polyelectrolyte brushes \cite{yu_multivalent_2018}. In the latter case, addition of minute amounts of trivalent ions to a solution of monovalent ions leads to a transition from a frictionless to a frictional state under load. The existence of such a transition in our systems should lead to the observation of shear-thickening.

\begin{figure} 
\centering
\includegraphics[width=\linewidth]{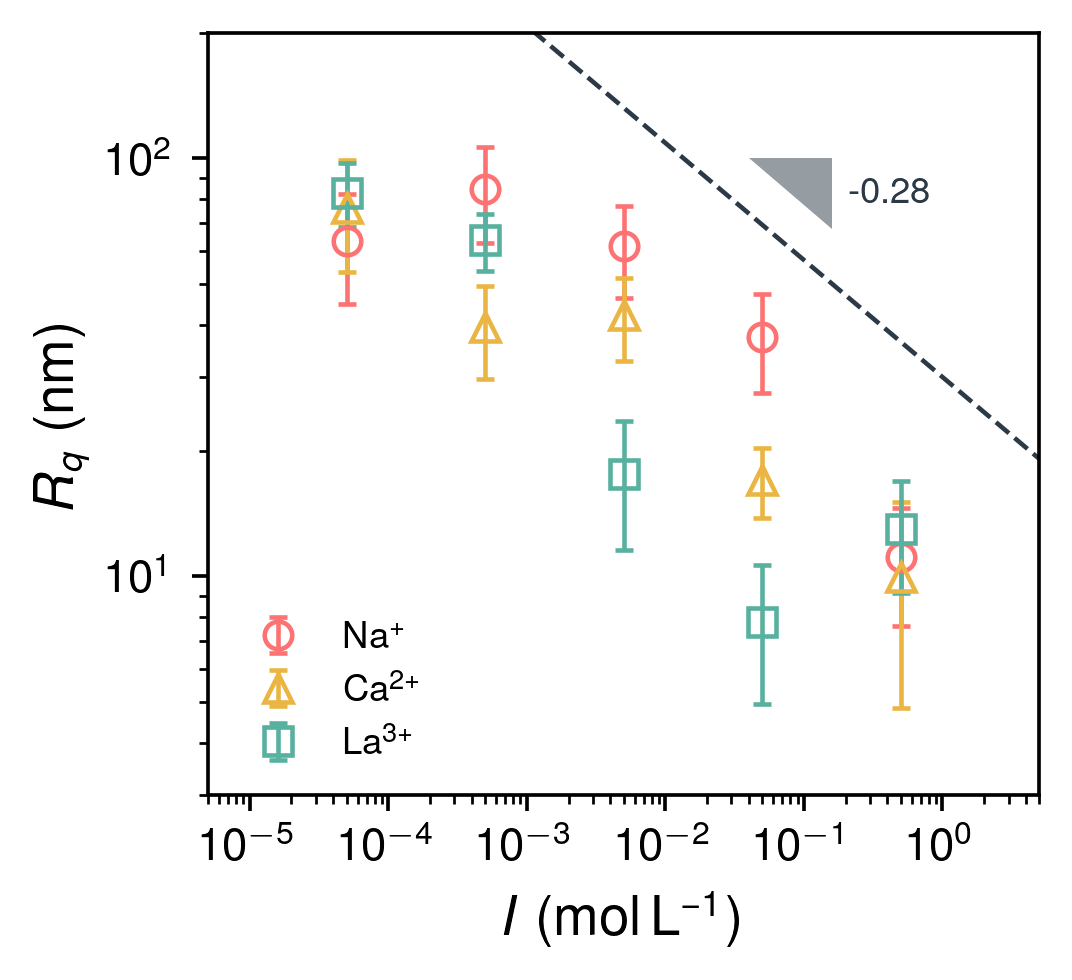}
\caption{Impact of ionic strength and ion valency on the surface roughness $R_q$ of HP60 particles in the presence monovalent cations Na$^+$ (\textcolor{peach}{$\bigcirc$}), divalent cations Ca$^{2+}$ (\textcolor{honey}{$\triangle$}), and trivalent cations La$^{3+}$ (\textcolor{keppel}{$\square$}). The values were obtained from a series of AFM images performed on the apex of one HP60 particle per type of salt. The dashed grey line corresponds to a scaling law with an exponent of $\alpha \approx -0.28$ for Na$+$ and Ca$^{2+}$.}
\label{fig-9}
\end{figure}

Second, Figure \ref{fig-9} displays the evolution of the particle roughness $R_q$ as a function of the ionic strength $I$ for mono- and multivalent salts. \textcolor{black}{As the salt concentration increases, we observed that the avalanche angle also increases and is correlated to the screening of the eletrosteric repulsion and phase separation occurring between the polymer network PMAA-Na of HP60 particles and the solvent. However, we see that the roughness decreases above a threshold in ionic strength due to osmotic effects stemming from the addition of mono- of multivalent cations. This decrease in roughness $R_q$ follows} a power law with an exponent -0.28, at least in the case of NaCl and CaCl$_2$. This exponent is similar to the one of the power law characterizing the decrease of the thickness of polymer brushes in the presence of salt.\cite{pincus_colloid_1991, ross_polyelectrolyte_1992} The connection between the roughness of the particles and the thickness of a polyelectrolyte brush is far from obvious, especially since there is no reason that the surface layer of our particles be made of a brush. The similarity in the exponent calls into question the structure of our particles. It would be interesting to probe in more detail the nature of the surface layer of our particles, how far it extends in the bulk, and to compare its features to those of a brush. Also, our protocol makes it possible to produce core-shell particles where a thin layer of PMAA-Na sits on a PMMA core, and it would be interesting to have a look at these systems with an AFM to probe further this resemblance with a brush.


HP60 particles open great perspectives for rheological studies of dense granular suspensions. Preliminary experiments with a pressure-imposed rheometer, the Capillarytron,\cite{etcheverry_capillary-stress_2023} suggest that suspensions of these particles remain frictionless over the whole range of shear stresses and pressures accessible to the apparatus, suggesting that the pressure applied to the particles never overcomes the repulsion force. Current work in our group investigates the consequences of these preliminary results.

\section*{Author Contributions}
E. G., M. R. and N. S. acquired the funding supporting this research; L. B., E. G., M. R. and N. S. conceptualized research; L. B. carried out the experiments; B. B., N. S. and L. B. carried out the AFM measurements; L. O. designed the rotating drum cell; E. G., M. R. and N. S. supervised the work; L. B., E. G., M. R. and N. S. analyzed the data; L. B. wrote the initial draft of the paper; L. B., E. G., M. R. and N. S. reviewed and edited the paper.

\section*{Conflicts of interest}
There are no conflicts to declare.

\begin{acknowledgments}
This work was supported by the project 80Prime CNRS 2020 ``Suspensions mod\`eles de particules fonctionnalis\'ees (SUMO)''. We thank Yoël Forterre and Bloen Metzger for their kind help during the preliminary measurements with their pressure-imposed rheometer. 
\end{acknowledgments}


\balance


\bibliography{references} 
\bibliographystyle{apsrev} 

\end{document}